# Information Technology Platforms: Conceptualisation and a Review of Emerging Research in IS Research

**Ruonan Sun**
Research School of Management
The Australian National University
Canberra, Australia
Email: ruonan.sun@anu.edu.au

**Shirley Gregor**
Research School of Management
The Australian National University
Canberra, Australia
Email: shirley.gregor@anu.edu.au

**Byron Keating**
Research School of Management
The Australian National University
Canberra, Australia
Email: byron.keating@anu.edu.au

## Abstract

The concept of an information technology (IT) related "platform" is broad and covers phenomena ranging from the operating system Linux to the Internet. Such platforms are of increasing importance to innovation and value creation across many facets of industry and daily life. There is, however, a lack of common understanding in both research and industry about what is mean by the term "platform" when related to IT. This lack of consensus is detrimental to research and knowledge development. Thus, the aims of this study are to: (i) conceptualise an IT-platform by identifying its distinguishing dimensions and show the concept's extension to ecosystems; and (ii) identify important current research directions for the IT-platform concept. To achieve these aims a systematic literature review was undertaken with 132 relevant articles taken from major information systems journals, conferences, and business publications. The study contributes by providing a sound base for future research into IT-platforms.

## 1   Introduction and Background

Digital industry platforms are a technology trend that is seen as having a profound impact on enterprises. "Digital industry platforms are fuelling the next wave of breakthrough innovation and disruptive growth. Increasingly, platform-based companies are capturing more of the digital economy's opportunities for strong growth and profitability … platform-based ecosystems are the new plane of competition" (Accenture 2015 p. 50).

In research, the term "platform" has become increasingly prevalent during the past decade across a range of academic disciplines (e.g. Eisenmann et al. 2011; Boudreau and Jeppesen 2014; Evans 2009). Thomas et al (2014) reviewed management research and identified four streams of platform research based on elements including construct, unit of analysis, value creation, and value appropriation. Porch et al. (2015) extended Thomas et al.'s study by investigating platforms both intra-organisationally and inter-organisationally. However, both studies mention that focusing only on a management perspective is a limitation; and call for further studies to explore the boundaries of the platform concept beyond the management domain.

Information systems and information technology related fields (IS/IT) are an important and unique domain for the study of platforms because: (i) IT-platforms provide the foundations that enable a large family of applications and related business practices (Fichman 2004); and (ii) IT-platforms are shared by complementary goods that frequently interoperate with the core technology foundation to add functionality (Tiwana et al. 2010). For example, the platform *WeChat* (*WhatsApp*'s highly successfully competitor in China) enables a variety of complementary functions such as chatting, gaming, shopping, and banking that create significant business value by interacting with the platform and each other. IT can play a crucial role in establishing foundations and creating business opportunities for stakeholders from other functional areas.

To date IS research focusing specifically on platforms has been somewhat limited. Of the 132 articles identified in our examination of the IT-platform literature, fewer than half attempt to explicitly or implicitly define an IT-platform. Moreover, the review of IS literature reveals that there is an absence of a consistent understanding with respect to what IT-platforms are and why IT-platforms are worth investigating. The aims of this study, therefore, are to (i) provide a sound definition for the IT-platform concept by identifying its distinguishing dimensions and show the concept's extension to ecosystems; and (ii) identify important current research themes around the IT-platform concept. Our scope at this point is restricted to the concept of a single IT-platform. It is acknowledged that more complex situations exist when platforms overlap, but such situations are beyond our current study.

IT-platform has been applied in an extensible range of research streams, such as IT investment (Taudes et al. 2000), IT governance (Hagiu 2014), and IT performance (Banker et al. 2011). However despite a critical mass in IT-platform research, there has been no comprehensive understanding of IT-platform as a concept in the IS discipline. Comprehensive understanding of concepts is vital because concepts are seen as basic building blocks for theorising (Dubin 1969); and represent "mental configuration" of given phenomena (Bacharach 1989). Thus, the principle contributions of this study are twofold. First, we provide a foundation for further research by conceptualising the IT-platform concept. Second, we identify a framework of research themes and directions related to IT-platforms. Overall, this study methodologically reviews, analyses, and synthesises IT-platform related literature within the IS discipline, thereby offering a sound base for researchers and practitioners alike for further work.

The study is structured as follows. The next section outlines in detail the research design. The subsequent section presents an overview of findings from a systematic literature review followed by our conclusions.

## 2 Research Method

Our study employed a systematic review method to investigate the development of the IT-platform concept in the IS domain (Levy and Ellis 2006). A systematic literature review is a powerful tool to deal with a large number of literature sources and has been extensively employed in IS review articles (e.g. Fielt et al. 2014; Proch et al. 2015). While Boell and Cecez-Kecmanovic (2015) question the systematic review as a general method for literature reviews, they suggest that this type of review is particularly useful for pursuing relatively clear and straight-forward research purposes. Our purpose is relatively straight-forward and thus the choice of a systematic review is appropriate.

The subsequent analysis follows the recommendations by Okoli and Schabram (2010) and comprises: (1) planning for the review – specifying research purpose and protocol; (2) selecting literature from databases to identify relevant publications; (3) extracting retrieved publications to match the research purpose; and (4) executing the analysis and synthesising the findings on the basis of extracted articles.

### 2.1 Planning

The first step in any systematic review is a clear identification of the intended purpose of the review (Okoli and Schabram 2010; Boell and Cecez-Kecmanovic 2015). The purpose of this paper, as stated in Section 1, includes a comprehensive archival analysis of IS literature on IT-platforms. Furthermore, a prior protocol is a critical element in the process of conducting a high-quality literature review (Okoli and Schabram 2010). This study follows a protocol with a basic pre-codification schema, which captures what IT-platforms are and why researchers study IT-platforms. It addresses definitions and research themes in the examined IS literature. This planning step serves as a roadmap to support the review.

### 2.2 Selection

Identifying the literature sources is the main criteria for systematic selecting relevant articles (Webster and Watson 2002). As the purpose of this study is to investigate and synthesis IT-platforms research from an IS perspective, the focus is on the clusters of literature targeted in the IS community. Webster and Watson (2002, p. xv) suggest "a complete review covers relevant literature on the topic and is not confined to one research methodology, one set of journals, or one geographic region". Therefore, a sampling frame that includes the main IS outlets was developed, giving a list of journals and conferences. Two sets of IS articles were included in the review. In the first set, articles published in the eight journals listed by the Association of Information Systems (AIS) as the Senior Scholars' Basket of 8 Journals[1] were selected as they represent the top quality in the IS domain. Moreover, articles in the top 3 business publications[2], ranked by the Financial Times, were included to pursue completeness and to capture frontline practical ideas overtime. The three business publications are also highly ranked in IS journal rankings (e.g. ISWorld). In the second set, to ensure the literature review is as comprehensive as possible, articles in the proceedings of major IS conferences were also investigated. Five AIS affiliated conferences[3] were chosen.

## 2.3 Extraction

Following the research purpose and protocol, extraction was conducted in two rounds. First, "platform" was searched as the keyword in the title, abstract, and keywords of the targeted databases for the basket of 8 journals and IS conferences. Further, articles that use "IS platform", "IT platform", "digital platform", or "technology platform" in their titles were searched for in the top 3 business publications because such articles often do not have an abstract or keyword section; and the focus is on extracting articles where IT-platform is the central focus. Second, to ensure that only relevant IS research articles are reviewed, we excluded articles that use non-relevant dictionary meanings (e.g. Capability Maturity Model), discipline specific usage (e.g. online payment platforms), and methods-related usages (e.g. Regional Development Platform Method). As a result, 51 articles from the basket of 8 journals, 63 articles from IS conferences, and 18 articles from business publications were yielded (see Figure 1). It is notable that, although the platforms notion has been under investigation since 1997, the sample shows a relatively small portion of research attention from the pool of IS outlets but with a rapid rising trend.

Note that while a systematic approach is followed in extracting the most relevant articles for this review, there can be and will be some suitable articles that are excluded due to the searching approach and resource availability. Missing articles could occur in any review (Webster and Watson 2002), but this study has defined a feasible and appropriate scope to extract relevant articles that fit the research purpose (e.g. Bandara et al. 2009).

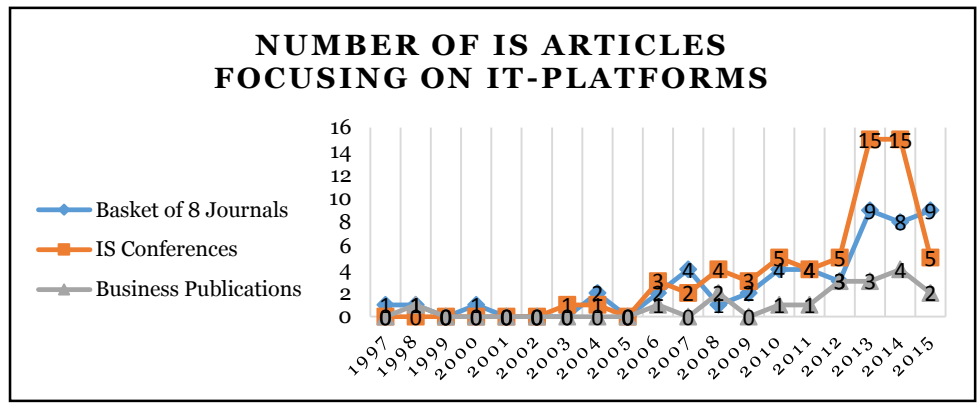

Figure 1. Number of IS articles focusing on IT-platforms (up to the 1st quarter 2015)

## 2.4 Execution

Our process of developing a definition for the IT-platform concept drew on advice from Eisenhardt (1989), Osigweh (1989); and Suddaby (2010). Following Suddaby (2010) we first examined prior definitions of platforms in the existing literature, to establish the historical lineage of the concept. Subsequently we used the process suggested by Eisenhardt (1989) of taking existing definitions and cases and preparing a table that categorizes and compares the attributes (dimensions) attributed to IT-platforms in prior work.

---

[1] See http://aisnet.org/general/custom.asp?page=SeniorScholarBasket for more details (retrieved 1 July 2015). The Senior Scholars' Basket of Journals are: European Journal of Information Systems, Information Systems Journal, Information Systems Research, Journal of AIS, Journal of Information Technology, Journal of MIS, Journal of Strategic Management Information Systems, and MIS Quarterly.
[2] See https://library.mcmaster.ca/find/ft-research-rank-journals for more details (retrieved 1 July 2015). The top 3 ranked business publications are: California Management Review (ranked#8), Harvard Business Review (ranked#12), and MIT Management Review (ranked#35).
[3] See http://aisnet.org/?page=Conferences for more details (retrieved 1 July 2015). The five AIS affiliated conferences are: International Conference on Information Systems (ICIS), Americas Conferences on Information Systems (AMCIS), Pacific Asia Conference on Information Systems (PACIS), European Conference on Information Systems (ECIS), and Australasian Conference on Information Systems (ACIS).

# 3 Review and Findings

This section discusses the systematic review of IT-platform focused articles from the IS literature. First, we introduced some key comments on the definition of concepts. Second, prior definitions of the concept were studied. Finally, a new definition of the IT-platform concept was developed based on the evidence and definitions in the sample articles. Third, key IT-platform research themes were specified.

## 3.1 Definition of Concepts

Concepts are seen as essential building blocks in theorising (e.g. Bacharach 1989; Dubin 1969) and the need for precise definition of concepts where possible is stressed (e.g. Osigweh 1989; Weber 2012). The terminology around "concepts" themselves is mixed. Some authorities refer to concepts and constructs as much the same thing (e.g. Whetten 1989) and in this essay we will not make a sharp differentiation between them. Further, the process of achieving clarity about what the concepts in our theorizing mean involves reflection on the underlying ontological and epistemological assumptions we believe apply to the situation at hand (see Weber 2012). For example, do the concepts refer to things that are socially constructed or to objects with an independent physical existence? We assume here that concepts can do either, with in-depth discussion of this point lying beyond the scope of the essay.

Different views on the definitions of concepts exist. The Cambridge Dictionary of Philosophy (1999) lists 18 different kinds of definition. Dubin (1969) in his work on theorizing prefers to use the term "units" rather than concepts for the basic building blocks of theory, with units being distinguished (defined) by the properties they possess. He uses the language of set theory to describe different types of units. For example, a unit (construct) can be an individual unit, or a member of a set (class), for which membership is defined in terms of having one or more common properties. Similar understanding can be found in authors such as Parsons and Wand (2013) in their treatment of classification principles and Osigweh (1989), who sees it as important in offering a definition to say what is not included in the definition (i.e., what does not belong to the defined set of things).

In addition, however, in many fields, it is not unusual to define what a complex thing is by saying what the thing is composed of: for example, in the definition of an atom. This usage is common in information systems: for example, the key concept of an "information system" itself is usually defined as consisting of a number of components (input, output, processing, and feedback), which bear a structural relationship to each other (e.g. see Stair and Reynolds 2012). Specification of such concepts is often aided by a diagrammatic representation. Bagozzi (1984) uses the term "structural definition" for something of this nature. An "IT-platform" may need to be defined in this way, as even at first glance the concept appears to refer to a complex thing with a number of component parts that are structurally related.

Pragmatically, for the process of defining a concept we heed the advice of Suddaby (2010, p. 2010) who says that good definitions should: (i) "capture the essential properties and characteristics of the phenomena under consideration"; (ii) "avoid tautology or circularity"; and (iii) "should be parsimonious". Suddaby also recommends showing the scope of the construct and relationships among constructs, including prior historical constructs on which a newer construct is based. Eisenhardt (1989, p. 542) describes the process of developing a construct definition from cases and notes that "many researchers rely on tables to summarize and tabulate the evidence underlying the construct" (citing Miles and Huberman 1984; Sutton and Callahan 1987).

## 3.2 Limitations in Understanding IT-Platforms

Although the concept "platform" is mentioned often in the IS literature, understanding on the precise meaning of the concept has not been achieved. The definitions of platforms vary widely between the general and the specific. For example, Gawer (2009, p. 2) more generally refers to platforms as "technological building blocks, providing an essential function to a technological system – which acts as a foundation upon which other firms can develop complementary products, technologies, or services". More specifically, for Donders et al. (2014, p. 88) a platform may refer to "a hardware configuration, an operating system, a software framework or any other common entity on which a number of associated components or services run".

There were 47 attempt to explicitly or implicitly define platforms among the 132 sample articles. Table 1 shows 15 selected sample definitions that reveal various understanding of IT-platforms. Based on the examined IS literature, IT-platform could be understood with different emphasis, for example, as a technology set on which complements can run (Fichman 2004; Meyer and Seloger 1998), a codebase for extension development (Tiwana 2015; Taude et al. 2000), a two- or multi-sided market enabled by technologies (Tan et al. 2015; Basole 2009); or technology and business infrastructure that enables enterprises activities (Richardson et al. 2014; Rai et al. 2006). However, there is limited understanding of the actual conceptualisation of IT-platforms in the literature so far. A comprehensive understanding of the IT-platform concept is vital because it is the foundation to building theories on IT-platforms (Suddaby 2000); and acts as an operational configuration of IT-platform related phenomena (Bacharach 1989).

| Sample Source | Definition |
|---|---|
| Taudes et al. 2000 | In general, a software platform is a software package that enables the realisation of application systems. |
| Richardson et al. 2014 | IT platforms enable a business infrastructure that shapes the capacity of firms to launch frequent and varied competitive actions, which results in improved performance. |
| Fichman 2004 | An IT-platform is broadly defined as a general-purpose technology that enables a family of applications and related business opportunities. |
| Tiwana 2015 | A platform refers to an extensible technological foundation and the interfaces used by extensions that interoperate with it. |
| Ceccagnoli et al. 2012 | Platforms are defined as the set of components used in common across a product family whose functionality can be extended by applications. |
| Rai et al. 2006 | In the supply chain management (SCM) context, an IT-platform enables consistent and real-time transfer of information between SCM related applications and functions that are distributed across partners. |
| Banker et al. 2011 | In the e-commerce context, an IT-platform is a website that allows participants to deposit margin money to ensure that they have resources to settle a dispute advice, and provides access to trading practices. |
| Markus and Loebbecke 2013 | Digital platform supporting simultaneous use by multiple companies, each of which can independently customise business process for its own ecosystem. |
| Tan et al. 2015 | The notion of platforms were defined as "two-sided markets", which refers to a market with two distinct sides that benefit from network effect by interacting on a common platform. |
| Shaw and Holland 2010 | The platform concept is used to label the structural level whose behaviour supports some higher level phenomena. |
| Giessmann and Stanoevska 2012 | Platforms and related components is a set of technology that is developed in emerging ecosystems of their-party developers. |
| Basole 2009 | Technology platforms are multi-sided markets since they bring together various types of participants or sides. |
| Heitkotter et al. 2012 | In the mobile context, platform refers to the symbolic combination of hardware, operating systems, and app store. |
| Saarikko et al. 2014 | The platform itself may be defined as a core of fixed set of attributes that can be extended by applications or complements to the benefit of adopters as well as backing firms. |
| Meyer and Seliger 1998 | Platform is a set of subsystems and interfaces that form a common structure from which a stream of derivative products can be efficiently developed and produced. |

*Table 1. Samples of Various IT-Platform Definitions*

To summarise, examining definitions of IT-platforms in the IS literature reveals that the existing IT-platform literature is characterised by diverse treatments of the concept; and lack of a systematic conceptualisation that captures in full the essential dimensions of the IT-platform concept. An explication and conceptualisation based on a systematic review is a necessary first step towards a more precise definition and knowledge advancement. The next section analyses in detail the dimensions of the IT-platform concept and develops a comprehensive definition. The definition leads to a conceptual model in diagrammatic form.

### 3.3 Conceptual Model and Definition of an IT-Platform

An examination of the IT-platform concept in the existing IS literature indicates that an IT-platform could be a technological foundation that allows complementary add-ons, or a two- or multi-sided market that facilitate exchanges, or some other representation. Our objective is to develop a comprehensive definition of an IT-platform and its key dimensions based on a systematic review. We begin by identifying candidate dimensions in an open coding process (Berg 1989) in the sample articles. These candidate dimensions were then grouped into categories where the candidates appeared to be employing different terminology for the same thing. The process continued until all candidates were grouped. The authors discussed areas of potential disagreement until consensus was reached. Table 2 depicts the core dimensions (categories) and alternative terminologies.

| Dimension | Examples of Alternative Terminologies |
|---|---|
| Technological base | • The basis of certain applications (Taudes et al. 2000)<br>• The set of components (Ceccagnoli et al. 2012; Rai et al. 2006)<br>• A general purpose technology (Fichman 2004; Song et al. 2013)<br>• An extensible codebase (Tiwana et al. 2010; Tiwana 2015)<br>• A core fixed set of attributes (Saarikko et al. 2014)<br>• Core infrastructure (Maurer and Tiwana 2012)<br>• Core products or services (Anderson Jr. et al. 2014)<br>• Common architecture (Bergvall-Karebron and Howcroft 2014)<br>• A common resource (Ghazawneh and Henfridsson 2010)<br>• A common structure (Giessmann and Stanoevska 2012)<br>• A common entity (Eurich et al. 2011)<br>• Technologies (Riemer and Richter 2010; Hagiu 2014) |
| Standards | • Standard (Rai et al. 2006)<br>• A set of rules (Kraemer et al. 2010)<br>• Interfaces (Meyer and Seliger 1998, Tiwana et al. 2010)<br>• Application programming interface (Lahiri et al. 2010)<br>• Meticulous platform interface (Tiwana 2015) |
| Add-ons | • Applications (Taudes et al. 2000; Ceccagnoli et al. 2012)<br>• Distributed applications (Rai et al. 2006)<br>• Complementary extensions- add-ins, modules, and apps (Tiwana et al. 2010)<br>• Complementary components (Bergvall-Kareborn and Howcroft 2014)<br>• Complementary products (Spagnoletti et al. 2015)<br>• Complementary models (Hilkert et al. 2011)<br>• Complementary innovations (Gawer and Gusumano 2008)<br>• Complementors (Suarez and Kirtely 2012)<br>• Complements (Saarikko et al. 2014)<br>• Associated components (Eurich et al. 2011)<br>• Subsystems (Meyer and Seliger 1998)<br>• Proprietary elements (Gawer and Gusumano 2008)<br>• Plug-ins (Jain et al. 2006) |
| Interoperability | • Interoperate (Tiwana et al. 2010)<br>• Interoperability (Tiwana 2015)<br>• Extend (Anderson Jr. et al. 2014)<br>• Connect (Koh et al. 2014; Suarez and Kirtley 2012)<br>• Real-time connectivity (Rai et al. 2006)<br>• Ways of connecting (Riemer and Richter 2010)<br>• Sharable (Dhar and Sundararajan 2007) |
| Transactionality | • Interactions (Hagiu 2014)<br>• Transactions (Mantena and Saha 2012)<br>• Exchanges (Avgerou and Li 2013) |
| Governance | • Coordination (Tiwana et al. 2010; Saarikko 2015)<br>• Platform governance (Tiwana 2015)<br>• Transparency – governance processes (Hilkert et al. 2011) |

*Table 2. Dimensions of the IT-Platform Concept*

### 3.3.1 Technological Base

A "technological base" refers to a technological foundation that is highly reusable and allows add-ons to be developed. Alternative terminologies in the literature include codebase (Tiwana et al. 2010), general purpose technology (Fichman 2004), common architecture (Bergvall-Karebron and Howcroft 2014), and basis of certain applications (Taudes et al. 2000). A technological base often encounters a stable-vs.-change trade-off that is – on the one hand, it must accommodate changes unforseen at the time the technological base was created; on the other hand, it must permit changes to individual add-ons without compromising its ability to function together again.

### 3.3.2 Standards

A "standard" refers to design rules that allow developers who access the platform at different times from different locations to make the same assumptions about the parts of the platform. Terminologies with similar meaning include interface (Meyer and Seliger 1998) and rule (Kraemer et al. 2010). Standardisation represents "the degree to which add-ons interact with the technological base using stable, well-documented, and redefined standards" (Tiwana et al. 2010 p. 679). To realise success from IT-platform standardisation, it is important to understand the trade-off between the stable value capture that a controlled standard yields and the rapid growth that an open standard facilitates (Dhar and Sundararajan 2007).

### 3.3.3 Add-Ons

An "add-on" refers to a software extension that connects to the platform technological base to add functionality. Researchers use terminologies such as application (Ceccagnoli et al. 2012), complement (Saarikko et al. 2014), modules (Tiwana et al. 2010), and subsystems (Meyer and Seliger 1998) to denote similar meanings. The addition and improvement of add-ons can enable capabilities and business models that would not exist otherwise. The value of an IT-platform thus depends on the add-ons that can be implemented (Taudes et al. 2000), e.g. apps in AppStore, games for PlayStation, and software for Windows.

### 3.3.4 Interoperability

"Interoperability" refers to the ability to interact between a technological base and add-ons at the technical level, such as an application programming interface (API) connection. Alternative terminologies with similar meaning are extend (Anderson Jr. et al. 2014), accessibility (Hilkert et al. 2011), ways of connecting (Riemer and Richter 2010) etc. Interoperability is an important antecedent that represents an IT-platform's ability to enable add-ons to contribute to the technological base.

### 3.3.5 Transactions

"Transactions" refers to interactions within an IT-platform ecosystem in ways that advance human interests at the non-technical level, such as e-marketing buying and selling[4]. Alternative terminologies such as interaction (Hagiu 2014) and exchange (Avgerou and Li 2013) are used in the literature to express similar meaning. The focus of IT-platform transactions is on platform-based business practices and exchanges. For example, IT-platform based business networks such as Taobao aggregate information content produced by exchanges between sellers and buyers, thereby creating value by harnessing transaction-generated data.

### 3.3.6 Governance

"Governance" broadly refers to policies, structures, processes, and mechanisms involved in managing an IT-platform. Some common IT-platform governance arrangements include relationship patterns, licencing agreements, and features to managing communication and exchange. The central notion of IT-platform governance is the tension between control by a platform owner and autonomy among independent developers and users (Tiwana et al. 2010). On the one hand, platform owners can increase profits through an optimal decision in terms of the degree of control to downstream add-ons. To promote subsequent innovations, a platform owner can choose either a closed contract that centralises platform profits or an open contract that stimulates greater add-on development depending on strategy. On the other hand, developers and users can choose to participate in either a closed platform or an open one depending on intellectual property and demands.

---

[4] An artificial trading agent buying and selling in a share market would still be initiating transactions as we are defining them, although there no human actor directly involved.

## 3.4 Ecosystem

Further, the definitions of IT-platforms are often expanded with the introduction of the concept of "ecosystem" (e.g. Hurni and Huber 2014; Anderson Jr., et al. 2014; Maurer and Tiwana 2012; Goldback and Kemper 2014). A platform-centric ecosystem is "the collection of the platform and the modules specific to that platform" (Tiwana et al. 2010, p. 675), in which the network of business processes and innovations make a platform more valuable (Ceccagnoli et al. 2012). The notion of a platform-centric ecosystem is particularly important with complex IT-platforms, such as iOS and Windows, as noted by Saarikko (2015):

> "The resulting structure can be leveraged for mass-customisation within a value chain or in a wider *ecosystem*. The latter is especially common in the IT-industry where short lifecycles force specialisation and product architectures must be able to accommodate a high degree of modularity."

## 3.5 Conceptual Model and Definition of an IT-Platform

Figure 2 depicts the IT-platform as a conceptual model, suggesting the interrelationship of its constituent dimensions. It is not unusual in IT and IS to depict high-level concepts in terms of inter-related constituent parts, as in the OSI, or **O**pen **S**ystem **I**nterconnection, model which defines a networking framework (Zimmermann 1980). IS by definition concerns systems, which involve sub-systems, interconnectivity, and information passing. Thus, a definition that includes the notion of layered dimensions appears appropriate. Appendix A shows examples of IT-platforms and dimensions.

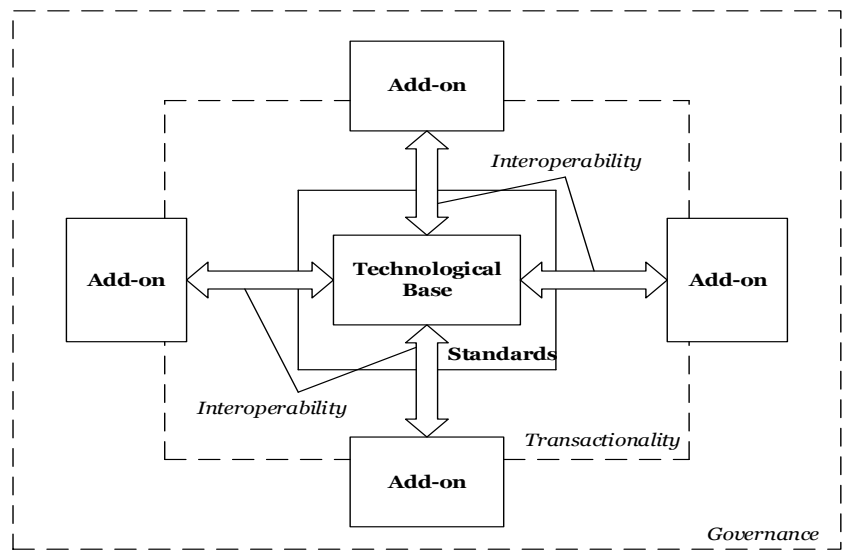

*Figure 2. IT-platform conceptual model*

Two theories serve as a theoretical lens to support the underlying logic of our IT-platform conceptual model. First, the general purpose technology (GPT) theory proposed by Lipsey et al. (2005, p. 98) argues that a GPT should have "scope for improvement and eventually comes to be widely used, to have many uses, and to have many spillover effects". A GPT refers to a single generic technology that is important enough to have a "protracted aggregated impact" (Jovanovic and Rousseau 2005; Lipsey et al. 2005). An important element for a GPT is the general purpose principle (GPP), as defined by Lipsey et al. (2005). The principle is employed by many different complementarities that are widely used around the GPT and across the economy. Such notions can help understand both the IT-platform constitution and interactions between key dimensions.

Second, modular systems theory suggests that complex systems (e.g. platform-centric ecosystems) comprise complementary goods that are always to some degree separated and combined; and the degree depend on "the 'rules' of the system architecture that enable (or prohibit) the mixing and matching of components" (Schilling 2000, p. 312). In our conceptual model, add-ons are the complementary subsystems that interact following defined standards and using stable interfaces within a platform-centric ecosystem, thereby increasing the value obtained by the core platform technological base (Tiwana et al. 2010). These arguments, in particular, support the design of IT-platform functionality in terms of how add-ons interoperate and transact with the platform core.

In summary, the conceptual model is the first step towards a systematic theory of IT-platforms. The model is developed through the lenses of GPT and modular systems. The conceptualisation leads to our definition of IT-platforms as follows:

*An IT-platform is defined as comprised of a technological base on which complementary add-ons can interoperate, following standards and allowing for transactions amongst stakeholders, within the platform-centric ecosystem.*

We believe this definition has general applicability, although at times people may focus on a particular dimension of the total ensemble.

## 4   Identifying Research Themes

We look at the objectives of published research to gain insights into research themes and directions using similar method as Thomas et al. (2014) and Fielt et al. (2014). The objective of a study is a critical part of research design. Identifying research objectives is important because a clear identified objective can explain why an article's results and contributions matter (Maxwell 2012). Researchers have raised questions about why IT-platforms thrive with (or without) commercial leadership and why we should care about platforms (Gawer 2009). The objectives of the sample articles were captured by carefully reading abstracts and introductory sections. A total of 128 objectives (explicit and implicit) were identified by searching and coding keywords and leading sentences[5] following an open coding process (Berg 1989).

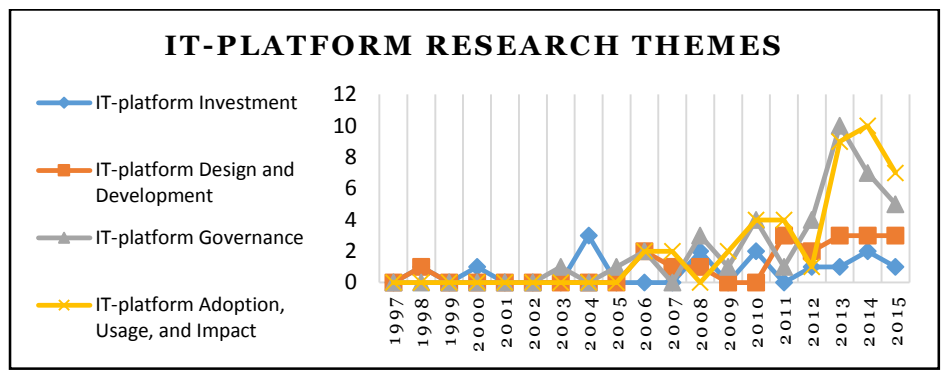

*Figure 3. Numbers of articles in different research themes*

We identify four main research themes, as presented in Figure 3 and Table 3, and show the correspondences to the dimensions of our conceptual model. The first theme looks at key determinants for platform investment and how investments contributes to platform value. Focal interests include platform owners' investment into technological base and developers' investment to complementary add-ons. The second theme studies enablers and inhibitors that influence the design and development of the technological base, add-ons, standards, and the ways they interoperate. The third theme focuses on IT-platform strategies and governance practices in terms of how to manage the interoperation and transactions between the technological base and add-ons. The fourth theme extends the scope of IT-platform governance to incorporate the ecosystem in investigating the adoption, usage, and impact of

---

[5] Keywords for identifying objectives are: objective, purpose, goal, and aim. Leading sentences for identifying objectives - such as "… this paper investigates an important factor in new product development …" (Anderson Jr. et al. 2014, p. 153)

IT-platforms. It is notable that IT-platform governance and adoption themes have been drawing increasing attention since 2011 (see Figure 3).

| Theme | Key Objective | Dimension |
|---|---|---|
| 1. IT-platform investment | To investigate and examine ways and factors that influence IT-platform investment decisions for stakeholders. | Tech. base, add-ons |
| 2. IT-platform design and development | To analyse factors that influence IT-platform design and development. | Tech. base, add-ons, standards, interoperability |
| 3. IT-platform governance | To explore governance practices and choices that serve to maintain a highly functional IT-platform. | Ecosystem |
| 4. IT-platform adoption, usage, and impact | To examine issues about IT-platform adoption, usage, and impact. | Ecosystem |

*Table 3. Framework for IT-Platforms Research Themes*

# 5 Limitations, Future Research, and Conclusion

Platforms are recognized as an increasing focus of innovation with the drive towards digital business (Accenture 2015; Yoo et al. 2010). An increasing number of IS research articles have paid attention to IT-platforms, with each bringing its own definition and theme to investigate or examine similar streams of issues. Unfortunately, a lack of integration has led to inconsistent understanding of IT-platform as a concept in the IS domain. This study examines the current understanding of IT-platforms as studied in the IS literature; and focuses on the definition and research directions for IT-platforms.

Although we have endeavoured to achieve the highest level of validity, this review is not without limitations. First, we have systematically selected and extracted articles from a range of sources (top IS journals, top business publications, and IS conferences) but the sample by necessity excludes many outlets. Second, the study strictly follows the review protocol and uses open coding process to manually conduct data analysis. However, the application of qualitative analysis tools (e.g. NVivo) could increase representation and mitigate the effect of biases (Leech and Onwuegbuzie 2007). Therefore, future research is expected to extend our analysis to second tier journals and use software packages to capture, code, and analyse sample literature.

As this study is still in-progress, we outline our plan for future work. First, findings of literature will be further synthesised to indicate how the new IT-platform conceptual model and definition will shape future research. Second, we propose to include industry stakeholders in IT-platform development and garner their perspectives. Third, a research agenda will be generated, through which researchers can more fully leverage the dimensions of IT-platforms when investigating the role of IT-platform in IS phenomena.

Despite the limitations, this review sheds light on our understanding of IT-platforms in the IS domain. This study (i) extracts six key dimensions of the IT-platform concept: technological base, standards, complementary add-ons, interoperability, transactionality, and governance; and (ii) identifies an important closely-related concept – a platform-centric ecosystem. We developed an integrative conceptual model that presents the interrelationships between the dimensions within the IT-platform ecosystem, contributing to a sound conceptualisation of IT-platforms. We also synthesise four main research themes. Future research directions are indicated by the growing interest in the last two themes.

This work is the first, to the best of our knowledge, to attempt to systematically and rigorously define the concept of an IT-platform and we hope that future research will find it a useful base for further refinement and extension.

# 6 References

Accenture. 2015. *Fjord Trends 2015 Report – How Digital Will Shape Consumer Expectations and Service Design.* DOI: http://www.accenture.com/SiteCollectionDocuments/us-en/accenture-fjord-trends-2015.pdf Retrieved 07 May 2015.

Anderson Jr., E. G., Parker, G. G., and Tan, B. 2014. "Platform Performance Investment in the Presence of Network Externalities," *Information Systems Research* (25:1), pp. 152-172.

Avgerou, C., and Li., B. 2013. "Relational and Institutional Embeddedness of Web-Enabled Entrepreneurial Networks: Case Studies of Netrepreneurs in China," *Information Systems Journal* (23), pp. 329-350.

Bacharach, S. B. 1989. "Organizational Theories: Some Criteria for Education," *The Academy of Management Review* (14:4), pp. 495-515.

Bagozzi, R. P. 1984. "A Prospectus for Theory Construction in Marketing," *Journal of Marketing* (48:1), pp. 11-29.

Bakos, Y., and Katsamakas, E. 2008. "Design and Ownership of Two-Sided Networks: Implications for Internet Platforms," *Journal of Management Information Systems* (25:2), pp. 171-202.

Bandara, W., Miskon, S., and Fielt, E. 2009. "A Systematic, Tool-Supported Method for Conducting Literature Review in Information Systems," *Proceedings of the 19th European Conference on Information Systems*, Helsinki, Finland.

Basole, R. 2009. "Structural Analysis and Visualization of Ecosystems: A Study of Mobile Device Platforms," *Proceedings of the 15th Conference on Information Systems*, San Francisco, US.

Berg, B. L. 1989. *Qualitative Research Methods for the Social Sciences*, MA: Allyn & Bacon.

Bergvall-Kareborn, B., and Howcroft, D. 2014. "Persistent Problems and Practices in Information Systems Development: A Study of Mobile Applications Development and Distribution," *Information Systems Journal* (24), pp. 425-444.

Boell, S. K., and Cecez-Kecmanovic, D. 2015. "On Being 'Systematic' in Literature Reviews in IS," *Journal of Information Technology* (30), pp. 161-173.

Boudreau, K. J., and Jeppesen, L. B 2014. "Unpaid Crowd Complementors: The Platform Network Effect Mirage," *Strategic Management Journal* (Online Version).

Donders, K., Pauwels, C., and Loisen, J. 2014. *The Palgrave Handbook of European Media Policy*, Palgrave Macmillan.

Ceccagnoli, M., Forman, C., Huang, P., and Wu, D. J. 2012. "Cocreation of Value in a Platform Ecosystem: The Case of Enterprise Software," *MIS Quarterly* (36:1), pp. 263-290.

Dhar, V., and Sundararajan, A. 2007. "Issues and Opinions – Information Technologies in Business: A Blueprint for Education and Research," *Information Systems Research* (18:2), pp. 125-141.

Dubin, R. 1969. *Theory Building*. New York: Free Press.

Eisenhardt, K. M. 1989. "Building Theories from the Case Study Research," *The Academy of Management Review* (14:4), pp. 532-550.

Eisenmann, T., Parker, G., and Van Alstyne, M. 2011. "Platform Envelopment," *Strategic Management Journal* (32:12), pp. 1270-1285.

Eurich, M., Giessmann, A., Mettler, T., and Stanoevska-Slabeva, K. 2011. "Revenue Streams of Cloud-based Platforms: Current State and Future Directions," *Proceedings of the 17th Americas Conference on Information Systems*, Detroit, US.

Evans, D. S. 2009. "How Catalysis Ignite: The Economics of Platform-Based Strat-ups", in Gawer, A. (ed.), *Platforms, Markets and Innovation*, Cheltenham, UK and Northampton, MA, US: Edward Elgar, pp. 99-128.

Fichman, R. G. 2004. "Real Options and IT Platform Adoption: Implications for Theory and Practice," *Information Systems Research* (15:2), pp. 132-154.

Fichman, R. G., Dos Santos, B. L., and Zheng, Z. 2014. "Digital Innovation as a Fundamental and Powerful Concept in the Information Systems Curriculum," *MIS Quarterly* (38:2), pp. 329-343.

Fielt, E., Bandara, W., Miskon, S., and Gable, G. 2014. "Exploring Shared Services from an IS Perspective: A Literature Review and Research Agenda," *Communications of the AIS* (34:54), pp. 1001-1040.

Gawer, A. 2009. "Platform Dynamics and Strategies: From Products to Services," in Gawer, A. (ed.), *Platforms, Markets and Innovation*, Cheltenham, UK and Northampton, MA, US: Edward Elgar, pp. 45-77.

Gawer, A. 2009. "Platforms, Markets and Innovation: An Introduction," in Gawer, A. (ed.), *Platforms, Markets and Innovation*, Cheltenham, UK and Northampton, MA, US: Edward Elgar, pp. 1-16.

Gawer, A., and Gusumano, M. A. "How Companies Become Platform Leaders," *MIT Sloan Management Review*, pp. 28-35.

Ghazawneh, A., and Henfridsson, O. 2013. "Balancing Platform Control and External Contribution in Third-Party Development: The Boundary Resources Model," *Information Systems Journal* (23), pp. 173-192.

Giessmann, A., and Stanoevska, K. 2012. "Platform as a Service – A Conjoint Study on Consumers' Perferences," *Proceedings of the 33rd International Conference on Information Systems*, Orlando, US.

Glaser, B. G. 2002. "Conceptualization: On Theory and Theorizing Using Grounded Theory," *International Journal of Qualitative Methods* (1:2), pp. 1-31.

Goel, L., Johnson, N. A., Junglas, I., and Ives, B. 2011. "From Space to Place: Predicting Users' Intentions to Return to Virtual Worlds," *MIS Quarterly* (35:3), pp. 749-771.

Goldbach, T., Kemper, V., and Benlian, A. 2014. "Mobile Application Quality and Platform Stickiness under Formal vs. Self-Control – Evidence from an Experimental Study," *Proceedings of the 35th International Conference on Information Systems*, Auckland, New Zealand.

Gorgan, J. L., and Gelinas Jr., U. J. 2007. "Managing the Internet Payment Platform Project," *Journal of IT* (22), pp. 410-419.

Hagiu, A. 2014. "Strategic Decisions for Multisided Platforms," *MIT Sloan Management Review*, DOI: http://sloanreview.mit.edu/article/strategic-decisions-for-multisided-platforms/ retrieved 2 July 2015.

Hilkert, D., Benlian, A., Hess, T. 2010. "Motivational Drivers to Develop Apps for Social Software-Platforms: The Example of Facebook," *Proceedings of the 16th Americas Conferences on Information Systems*, Lima, Peru.

Hilkert, D., Benlian, A., Sarstedt, M., and Hess, T. 2011. "Perceived Software Platform Openness: The Scale and its Impact on Developer Satisfaction," *Proceedings of the 32nd International Conferences on Information Systems*, Shanghai, China.

Jain, H., Rothenberger, M., and Sugumaran, V. 2006. "Flexible Software Component Design Using A Product Platform Approach," *Proceedings of the 27th International Conference on Information Systems*, Milwaukee, US.

Jain, R. P., and Ramesh, B. 2015. "The Roles of Contextual Elements in Post-Merger Common Platform Development: An Empirical Investigation," *European Journal of Information Systems* (24:2), pp. 159-177.

Jovanovic, B., and Rousseau, P. L. 2005. "Chapter 18 – General Purpose Technologies," P. Aghion and S. Durlauf (Eds.), *Handbook of Economic Growth*, North-Holland: Elsevier, pp. 1181-1224.

Keen, P. and Williams, R. 2013. "Value Architectures for Digital Business: Beyond the Business Model," *MIS Quarterly* (37:2), pp. 643-647.

Koh, T. K., and Fichman, M. 2014. "Multihoming Users' Perference for Two-Sided Exchange Networks," *MIS Quarterly* (38:4), pp. 966-977.

Kraemer, T., hinz, O., and Skiera, B. 2010. "Return on IT Investments in Two-Sided Markets," *Proceedings of the 21st Australasian Conference on Information Systems*, Brisbane, Australia.

Kuebel, H., and Zarnekow, R. 2014. "Evaluating Platform Business Models in the Telecommunications Industry via Framework-based Case Studies of Cloud and Smart Home Service Platforms," *Proceedings of the 20th Americas Conferences on Information Systems*, Savanah, US.

Lahiri, A., Dewan, R. M., and Freimer, M. 2010. "The Disruptive of Open Platforms on Markets for Wireless Services," *Journal of Management Information Systems* (27:3), pp. 81-110.

Leech, N. L., and Onwuegbuzie, A. J., 2007. "An Array of Qualitative Data Analysis Tools: A Call for Data Analysis Triangulation," *School Psychology Quarterly* (22:4), pp. 557-584.

Levy, Y., and Ellis, T. J. 2006. "A Systems Approach to Conduct an Effective Literature Review in Support of Information Systems Research," *Informing Science Journal* (9), pp. 181-212.

Lipesy, R. G., Carlaw, K. I., and Bekar, C. T. 2005. *Economic Transactions: General Purpose Technologies and Long Term Economic Growth*, US: Oxford.

Mantena, R., and Saha, R. L. 2012. "Co-opetition Between Differentiated Platforms in Two-Sided Market," *Journal of Management Information Systems* (29:2), pp. 109-140.

Markus, M. L., and Loebbecke, C. 2013. "Commoditized Digital Processes and Business Community Platforms: New Opportunities and Challenges for Digital Business Strategies," *MIS Quarterly* (37:2), pp. 649-653.

Maurer, C., and Tiwana, A. 2012. "Control in App Platforms: The Integration-Differentiation Paradox," *Proceedings of the 33rd International Conference on Information Systems*, Orlando, US.

Maxwell, J. A. 2012. *Qualitative Research Design: An Interactive Approach*, SEGA.

Mei, L., Khim-Yong, G., and Huseyin, C. 2014. "Investigating Developers' Entry to Mobile App Platforms: A Network Externality View," *Proceedings of the 22nd European Conference on Information Systems*, Tel Aviv, Israel.

Meiville, N., Kraemer, K., and Gurbaxano, V. 2004. "Review: Information Technology and Organizational Performance: An Integrative Model of IT Business Value," *MIS Quarterly* (28:2), pp. 283-322.

Meyer, M. H., and Seliger, R. 1998. "Product Platforms in Software Development," *MIT Sloan Management Review*, pp. 61-74.

Miles, M., and Huberman, A. M. 1984. *Qualitative Data Analysis*. Beverly Hills, CA: Sage Publications.

Mueller, J., Hutter, K., Fueller, J., and Matzler, K. "Virtual Worlds as Knowledge Management Platform – A Practice-Perspective," *Information Systems Journal* (21), pp. 479-501.

Okoli, C., and Schabram, K. 2010. "A Guide to Conducting a Systematic Literature Review of Information Systems Research," *Sprouts: Working Papers on Information Systems*, (10:2), DOI: http://sprouts.aisnet.org./10-26 Retrieved 30 June 2015.

Ondrus, J., Gannamaneni, A., and Lyytinen, K. 2015. "The Impact of Openness on the Market Potential of Multi-Sided Platforms: A Case Study of Mobile Payment Platforms," *Journal of IT*, DOI: 10.1057/jit.2015.7 retrieved 7 June 2015.

Osigweh, C. A. B. YG. 1989. "Concept Fallibility in Organizational Science," *Academy of Management Review* (14:4), pp. 579-594.

Parsons, J., and Wand, Y. 2013. "Extending Classification Principles from Information Modelling to Other Disciplines," *Journal of the Association for Information Systems* (5:14).

Porch, C., Timbrell, G., and Rosemann, M. 2015. "Platforms: A Systematic Review of the Literature Using Algorithmic Historiography," *Proceedings of the 23rd European Conference on Information Systems*, Munster, Germany.

Rai, A., Patnayakuni, R., and Seth, N. 2006. "Firm Performance Impact of Digitally Enabled Supply Chain Integration Capabilities," *MIS Quarterly* (30:2), pp. 225-246.

Riemer, K., and Richter, A. 2010. "Social Software: Agents for Change or Platforms for Social Reproduction? A Case Study on Enterprise Microblogging," *Proceedings of the 21st Australasian Conference on Information Systems*, Brisbane, Australia.

Rochet, J. C., and Tirole, J. 2006. "Two-Sided Markets: A Progress Report," *RAND Journal of Economics* (37:3), pp. 645-667.

Saarikko, T. 2015. "Digital Platform Development: A Service-Oriented Perspective," *Proceedings of the 23rd European Conferences on Information Systems*, Minster, Germany.

Saarikko, T., Jonsson, K., and Burstrom, T. 2014. "Towards an Understanding of Entrepreneurial Alertness in the Formation of Platform Ecosystem," *Proceedings of the 22nd European Conference on Information Systems*, Tel Aviv, Israel.

Schilling, M. A. 2000. "Toward a General Modular Systems Theory and Its Application to Interfirm Product Modularity," *Academy of Management Review* (25:2), pp. 312-334.

Schilpzand, P., Hekman, D. R., and Mitchell, T. R. 2015. "An Inductivity Generated Typology and Process Model of Workplace Courage," *Organization Science* (26:1), pp. 52-77.

Scholten, U., Janiesch, C., and Rosenkranz, C. 2013. "Inciting Networks Effects through Platform Authority: A Design Theory for Service Platforms," *Proceedings of the 34th International Conference on Information Systems*, Milan, Italy.

Schwarz, A., and Hirschheim, R. 2003. "An Extended Platform Logic Perspective of IT Governance: Managing Perceptions and Activities of IT," *Journal of Strategic Information Systems* (12), pp. 129-166.

Shaw, D. R., and Holland, C. P. 2010. "Strategy, Networks and Systems in the Global Translation Services Market," *Journal of Strategic Information Systems* (19), pp. 242-256.

Song, J., Baker, J., Choi, H., and Bhattacherjee, A. 2013. "Towards a Theory of Information Technology Platform Adoption," *Proceedings of the 34th International Conference on Information Systems*, Milan, Italy.

Spagnoletti, P., Resca, A., and Lee. G. 2015. "A Design Theory for Digital Platforms Supporting Online Communities: A Multiple Case Study," *Journal of IT* pp. 1-17.

Stair, R., and Reynolds, G. 2012. *Fundamentals of Information Systems*, Course Technology: CENGAGE Learning.

Suarez. F. F., and Kirtley, J. "Dethroning an Established Platform," *MIT Sloan Management Review*, pp. 35-41.

Suddaby, R. 2010. "Editor's Comments: Construct Clarity in Theories of Management and Organization," *The Academy of Management Review* (35:3), pp. 346-357.

Sutton, R., and Callaham, A. 1987. "The Stigma of Bankruptcy: Spoiled Organizational Image and Its Management," *Academy of Management Journal* (30), pp405-436.

Sviokla, J., and Paoni, A. J. 2005. "Every Product's a Platform," *Harvard Business Review*, DOI: https://hbr.org/2005/10/every-products-a-platform Retrieved 30 June 2015.

Taudes, A. 1998. "Software Growth Options," *Journal of Management Information Systems* (15:1), pp. 165-185.

Taudes, A., Feurstein, M., and Mild, A., 2000. "Options Analysis of Software Platform Decisions: A Case Study," *MIS Quarterly* (24:2), pp. 227-243.

Tan, B., Lu, X., Pan, S. L., and Huang, L. 2015. "The Role of IS Capabilities in the Development of Multi-Sided Platforms: The Digital Ecosystem Strategy of Alibaba.com," *Journal of the Association for Information Systems* (16:4), pp. 248-280.

The Cambridge Dictionary of Philosophy, 1999. Audi, R (Ed.), Cambridge University Press.

Thomas, L. D.W., Autio, E., and Gann, D. M. 2014. "Architectural Leverage: Putting Platforms in Context," *The Academy of Management Perspectives* (28:2), pp. 198-219.

Tiwana, A. 2015. "Evolutionary Competition in Platform Ecosystems," *Information Systems Research* (26:2), pp. 266-281.

Tiwana, A., Konsynski, B., and Bush, A. A. 2010. "Research Commentary – Platform Evolution: Coevolution of Platform Architecture, Governance, and Environmental Dynamics," *Information Systems Research* (21:4), pp. 675-687.

Wagelaar, D., and Van Der Straeten, R. 2007. "Platform Ontologies for the Model-Driven Architecture," *European Journal of Information Systems* (16), pp. 362-373.

Walravens, N. 2013. "The City as a Service Platform: A Typology of City Platform Roles in Mobile Service Provision," *Proceedings of the 19th Americas Conferences on Information Systems*, Chicago, US.

Weber, R. 2012. "Evaluating and Developing Theories in the Information Systems Discipline," *Journal of the Association for Information Systems* (13:1).

Webster, J., and Watson, R. T. 2002. "Analyzing the Past to Prepare for the Future: Writing a Literature Review," *MIS Quarterly* (26:2), pp. xiii-xxiii.

Whetten, D. A. 1989. "What Constitutes a Theoretical Contribution?," *The Academy of Management Review* (14:4), pp. 490-495.

Yang, L., Su, G., and Yuan, H. 2012. "Design Principles of Integrated Information Platform for Emergency Responses: The Case of 2008 Beijing Olympic Games," *Information Systems Research* (23:3-part-1), pp. 761-786.

Yoo, Y., Boland Jr., R. J., Lyytinen, K., and Majchrzak, A. "Organizing for Innovation in the Digitized World," *Organization Science* (23:5), pp. 1398-1408.

Zimmermann, H. 1980. "OSI Reference Model – The ISO Model of Architecture for Open Systems Interconnection," *IEEE Transactions on Communications* (28:4), pp. 425-432.

**Appendices**

**Appendix A: Examples of IT-Platform Ecosystems and Dimensions**

|  |  | IT-Platform Ecosystem |  |  |  |  |
|---|---|---|---|---|---|---|
|  |  | Linux | iOS | SAP | PlayStation | Taobao |
| **Dimension** | Technological base | Linux kernel | iOS core | SAP S/4HANA | Accelerated processing unit | Quad-Core Intel Xeon processor tech |
|  | Standards | Linux standard base | System View Controller | SAP Application Performance Standard | Move.Me Network Protocol | National unified payment gateway |
|  | Interoperability | Shell (user interface) – command-line interface (CLI) or graphical user interface (GUI) to achieve the best workflow for tasks. | iOS SDK with tools for developers to develop, install, run, and test add-ons, using Objective-C language and run on iOS base. | SAP cloud applications studio which enables partners to adapt and enhance the solution capabilities of SAP's Cloud solutions. | Cell Broadband Engine Architecture with Cell/B.E technology for game developing. | Taobao open platform API includes a suite of services covering 15 categories for app development. |
|  | Add-ons | Software in the Ubuntu Centre | Apps in the AppStore | SAP packages | Games, movies, apps et al. | Online chatting, reputation ranking |
|  | Transactions | Software service operations, e.g. sending and receiving messages | App downloading, service purchasing, app evaluating etc. | Information and value exchange through SAP ERPs and business suits | Game purchasing, media sharing, live broadcasting, play sharing etc. | Online chatting, online value transfer etc. |
|  | Governance | Free software licensing agreement | Apple developer licensing agreement | Quota arrangements, global partner network | PlayStation partner registration and agreement | Taobao platform service agreement, reputation management |

*Table A-1. Examples of IT-Platform and Dimensions*